\documentclass[%
reprint,
superscriptaddress,
 amsmath,amssymb,
 aps,
prb,
]{revtex4-2}

\usepackage{xcolor}
\definecolor{royalblue}{RGB}{65,105,225}
\definecolor{navy}{RGB}{0,0,128}
\definecolor{teal}{RGB}{0,128,128}
\definecolor{tomato}{RGB}{255,99,71}

\usepackage{graphicx}
\usepackage{dcolumn}
\usepackage{bm}
\usepackage[colorlinks, urlcolor=blue,linkcolor=blue,anchorcolor=blue,citecolor=blue]{hyperref}
\usepackage{subfigure}
\usepackage{multirow}

\begin{document}

\preprint{APS/123-QED}

\title{Spinon quantum spin Hall state in the kagome antiferromagnet with a Dzyaloshinskii-Moriya interaction}

\author{Li-Wei He}
\affiliation{%
 National Laboratory of Solid State Microstructures and School of Physics, Nanjing University, Nanjing 210093, China
}%

\author{Jian-Xin Li}%
\email{jxli@nju.edu.cn}
\affiliation{%
 National Laboratory of Solid State Microstructures and School of Physics, Nanjing University, Nanjing 210093, China
}%
\affiliation{Collaborative Innovation Center of Advanced Microstructures, Nanjing University, Nanjing 210093, China}

\begin{abstract}
We investigate the spin-$\frac{1}{2}$ antiferromagnetic Heisenberg model with a Dzyaloshinskii-Moriya interaction on kagome lattice, making use of the variational Monte Carlo technique. An exotic quantum spin state is found to arise from a melting of the $\boldsymbol{Q} = 0$ long-range magnetic order by a topological transition,  when a small anisotropic third nearest-neighbor antiferromagnetic Heisenberg interaction is turned on. This novel state is a gapped quantum spin liquid, characterized by a topological order with ground-state degeneracy $n_g = 4$ and topological entanglement entropy $\gamma = \ln 2$, suggesting it is an Abelian topological phase. Furthermore, the Chern numbers of the spin-up (-down) spinon occupied bands of this state are $C_{\uparrow \downarrow} = \pm 1$, respectively. From this perspective, this state is also a time-reversal symmetric (total Chern number $C_{total} = 0$) topological insulator with spinons as the chiral edge states, which carry opposite spin and move in the opposite direction. It is analogous to the quantum spin Hall state but the spin current is carried by deconfined spinons in a quantum spin liquid, so is dubbed as the spinon quantum spin Hall state.  
\end{abstract}

\maketitle


\section{\label{sec:introduction}introduction}

One of the exotic and intriguing phase receiving extensive attention and research in condensed matter physics is the quantum spin liquid (QSL)~\cite{Anderson-RVB-1973}, which is magnetic disordered and has the fractionalized elementary excitation called spinon with spin-$\frac{1}{2}$. One class of this phase is the gapped QSL, which is a crucial representative of the topological orders with the long-range many-body quantum entanglement. Among the gapped QSLs, the chiral spin liquid (CSL)~\cite{Wen-csl-1989}  breaks the time-reversal (TR)  symmetry  and usually has nontrivial Chern number to characterize itself. In a CSL, the spinons with up and down spins usually couple the same gauge field and have the same Chern number. For those gapped QSLs with TR symmetry, the ground-state degeneracy (GSD)~\cite{Wen-gsd-1989, Wen-gsd-1990} other than the Chern number is a good quantity to describe the global topological property. Furthermore, another important quantity, the topological entanglement entropy (TEE)~\cite{tee-levin-wen-2006-prl, tee-kitaev-2006-prl} related to the quantum dimension of topological excitaitons, is very useful for both chiral and achiral topological orders. 

Another exotic and extensively studied phase is the quantum spin Hall (QSH) state~\cite{qsh-kane-prl-2005, qsh-bernevig-prl-2006},  which is realized in  the  topological insulator~\cite{topo_insulator-rev-qi-2011} characterized by an insulating bulk gap and gapless edge or surface states (the bulk-edge correspondence of topological system) protected by TR symmetry. In the QSH state, the electrons with opposite spins move along the opposite direction on a given edge~\cite{qsh-kane-prl-2005, qsh-wu-prl-2006, qsh-xu-prb-2006}. Consequently, the two states on the edge possess the spin Chern number with opposite signs so that the QSH state is characterized by a Z$_\mathrm{2}$ topological number. 

Pictorially, both the gapped QSL with TR symmetry and the QSH state are insulators and topological nontrivial phases. However, the conventional topological insulator is a non (or weak)-interacting single particle state whose quasiparticles are electrons in the framework of Landau Fermi liquid. While, the QSL is a strong interacting Mott insulator without conventional Landau quasiparticles but the fractionalized excitations such as spinons. Hence, thus far, these two states are studied separatively from each other. Thus, an issue arises as to if there exists a possible exotic quantum state which is a gapped QSL with TR symmetry meanwhile the quasiparticles (spinons) exhibit similar QSH effect.

To search for QSLs, the kagome antiferromagnet is an appealing platform~\cite{Depenbrock-prl-z2, Ran-dsl-2007, Iqbal-prb-j1j2-dsl, He-prx-dmrg-dsl, j1j2-prb-2011-vmc-slaveboson, j12d-prb-phase-diagram-2015, dm-prb-2017-phase-diagram, j12d-prb-2015-csl, hu-j12d_ch-prb-2015-csl, xxz-csl-prl-2014, xxz-csl-prl-2015, j12d_jch-prb-2015-csl, PhysRevLett.101.117203, doi:10.1126/science.1201080,PhysRevLett.118.137202},
in view of its strong geometric frustration and the resulting strong quantum fluctuations.  Many novel many-body quantum states have been explored in this lattice recently ~\cite{PhysRevB.80.113102,PhysRevB.82.075125,PhysRevLett.106.236802,PhysRevB.85.144402,nature-kagome-2018-1,nature-kagome-2018-2,doi:10.1126/science.aav2873,PhysRevLett.125.247002,nature-kagome-2021-sc,nature-kagome-2022-sc-1,nature-kagome-2022-sc-2,nature-kagome-2022-sc-3}.
It is generally believed that the Heisenberg model with the nearest-neighbor (NN) antiferromagnetic (AFM) interaction $J_1$ on a kagome lattice hosts a QSL ground state.  However, it is controversial that this QSL is gapped~\cite{Depenbrock-prl-z2} or not~\cite{Ran-dsl-2007, Iqbal-prb-j1j2-dsl, He-prx-dmrg-dsl}. Further, the long-range AFM Heisenberg spin interactions beyond the NN term are always possible and will affect the properties of ground states. It is reported that the second NN AFM interaction $J_2$ favors a $\boldsymbol{Q} = 0$ magnetic order~\cite{j1j2-prb-2011-vmc-slaveboson, Iqbal-prb-j1j2-dsl, j12d-prb-phase-diagram-2015, dm-prb-2017-phase-diagram}.
However, the interplay between the $J_2$ term and one of the third NN AFM couplings which is across the diagonals of hexagons $J_d$ can induce a CSL~\cite{j12d-prb-2015-csl, hu-j12d_ch-prb-2015-csl, j12d-prb-phase-diagram-2015} with spontaneous TR symmetry breaking. And even, the CSL can arise in the XXZ model with anisotropic $J_2$ and $J_d$ terms~\cite{xxz-csl-prl-2014, xxz-csl-prl-2015}. 
These results point to the dominant role that a relatively large $J_d$ might play to stabilize the CSL though it alone does not. Moreover, the scalar three-spin interaction $\chi = \boldsymbol{S}_i \cdot \boldsymbol{S}_j \times \boldsymbol{S}_k$ (subscripts mean vertexes of each triangle), which breaks TR symmetry,  can naturally stabilize the CSL~\cite{hu-j12d_ch-prb-2015-csl, j12d_jch-prb-2015-csl}. Therefore, it suggests that the novel gapped QSL with TR symmetry we look for does not exist in these models.  

Another choice is to consider the Dzyaloshinskii-Moriya interaction (DM) interaction, whose effects in a kagome lattice  have recently been investigated theoretically and experimentally~\cite{dm-prb-2002, dm-prb-2015-csl, dm-prb-2017-phase-diagram, dm-2017-prl-slave-boson-csl, dm-prb-2017-thermal-hall-magnon,dm-prl-2008-experiment, dm-experiment-prb-2011, Lee-dm-kagome-2013, kagome_tee_np}. However,  it has been shown that this small interaction in fact favors the in-plane $\boldsymbol{Q} = 0$ magnetic order, when the direction of DM vector is perpendicular to the $xy$ plane and leads to the local vector chirality $\boldsymbol{\chi}_\mathrm{v} = \boldsymbol{S}_1 \times \boldsymbol{S}_2 + \boldsymbol{S}_2 \times \boldsymbol{S}_3 + \boldsymbol{S}_3 \times \boldsymbol{S}_1$.  Hence, if proceeding along this direction, we need to consider additional exchange interactions to melt this magnetic order.  We notice that in this magnetic order spins aline ferromagnetically along the diagonal direction of the hexagon in the kagome lattice. So, it arises a possibility to consider the interplay between the DM interaction and the third NN AFM interaction along the diagonal direction.

Based on these considerations, in this work, we study the nature of the quantum spin states in the $J_1$-$J_2$-$J_3$ kagome antiferromagnet with additional DM interaction and the third NN AFM interaction $J_d$ along the diagonal direction.  We do not intend to obtain the global phase diagram with so many spin interaction parameters, instead mainly focus on the study of quantum spin states emerging out of the $\boldsymbol{Q} = 0$ magnetic ordered state.  
When only $J_1$ term exists, our numerical simulation suggests a Dirac spin liquid (DSL) as the ground state, and the introduction of a weak DM interaction transits the system into the long-range $\boldsymbol{Q} = 0$ magnetic order. These results are consistent with previous researches~\cite{Ran-dsl-2007, Iqbal-prb-j1j2-dsl, He-prx-dmrg-dsl,dm-prb-2017-phase-diagram,dm-2017-prl-slave-boson-csl}. Therefore, without loss of generality, we will fix the magnitude of DM interaction $D$ as $0 \alt D \alt 0.2$ in our main calculations to investigate the effects of other longer-range AFM Heisenberg interactions. When a weak diagonal third NN interaction $J_d$ turns on, the magnitude $M$ of the magnetic moment for the $\boldsymbol{Q} = 0$ state decreases and eventually drops to zero when $J_d \simeq 0.21$ (setting $J_1=1$).  Our calculation suggests there is a continuous phase transition into a disordered phase, a gapped QSL with TR symmetry. We also show that when a small additional $J_2$ is present, this transition still occurs, for example a slightly larger $J_d \simeq 0.27$ is required with $J_2=0.05$ because the $J_2$ term favors the $\boldsymbol{Q} = 0$ order, as mentioned above. Obviously, this state can survive in a broad range of $J_d$. It is verified that the CSL and the so-called $cuboc1$ magnetic state breaking TR symmetry are not found in the range of $J_d$ we considered, while the $cuboc1$ order is indeed found in the large $J_d$ range, such as $D = 0.1, J_{2,3} = 0, J_d = 0.3$. Beyond the diagonal third NN term $J_d$, we have also checked the effects of the usual NN term $J_3$ and find that it further enhances the effect of suppressing the $\boldsymbol{Q} = 0$ order. Therefore, this phase transition is general in the sense of the reasonable DM interaction and diagonal third NN term, and the resulting gapped QSL with TR symmetry is robust. 

To describe the intrinsic property of the gapped QSL, we calculate the TEE and find $\gamma \simeq 0.748 $ on a torus, which agrees numerically with the exact value $\gamma_\mathrm{ideal} = \ln 2 \simeq 0.693$.  So, the quantum dimension $D_q=2$ is obtained via $\gamma = \ln D_q$. In the meantime, we find that its GSD is $n_g = 4$.  
These results suggest that the gapped QSL holds an Abelian topological order with $n_g=D_q^{2}$.
It is expected that the total Chern number of this QSL is zero because of TR symmetry. Interestingly, we find that the spin-up and -down spinons see opposite gauge fluxes and there is no coupling between them. As a result, we can independently define the Chern numbers for the spin-up and -down spinons, respectively. It turns out that the Chern number of the spin-up (-down) spinons is $C_{\uparrow(\downarrow)} = +(-)1$, which is right the Z$_2$ index~\cite{topo_chern_num-2017-review}. So, the spinons with different spins move along opposite directions on a given edge with the opposite chiral central charges $c_\pm = \pm 1/2$. That shows that this disordered state is an exotic gapped QSL with TR symmetry and shares the same properties as those of a QSH state at the mean time. Thus, we name it the spinon quantum spin Hall (SQSH) state. According to the topological long-range entanglement, ground-state degeneracy and the Chern number of this SQSH state, we suggest that it is a double-semion topological order (doubled Chern-Simons state), which is described by a sum of two topological quantum field theories with opposite chiralities and is suggested in the string-net model~\cite{PhysRevB.71.045110}. Generalizing to spin-$k$ ($k$ is integer) antiferromagnets, the spin liquids with parity and time-reversal symmetry have been proposed, such as the doubled CSL termed as CSL$_{+}$CSL$_{-}$ with $k=1$~\cite{PhysRevB.84.140404}, which holds the same topological properties with the SQSH state. In addition, this state has also been studied by the wire deconstructionism of the long-range entanglement topological phase~\cite{PhysRevB.90.205101}.

\section{\label{sec:model-method}model and method}

We start with the following model,
\begin{equation}
	\begin{aligned}
		H =&\  \sum_{\left( i,j \right)} J_{ij} \boldsymbol{S}_i \cdot \boldsymbol{S}_j
		+ D\sum_{\langle i,j \rangle}\boldsymbol{D}_{ij} \cdot \boldsymbol{S}_i \times \boldsymbol{S}_j.
	\end{aligned}
	\label{eq:hamiltonian}
\end{equation}
The first term in model (\ref{eq:hamiltonian}) is the AFM Heisenberg spin interaction, including the first, second and two anisotropic third NN spin couplings, expressed respectively as the $J_1$, $J_2$, $J_3$ and $J_d$ terms.  The second term denotes the DM interaction connecting the first NN bond spins, and the direction of the DM vector $\boldsymbol{D}_{ij}$ is perpendicular to the NN bond $\langle i,j \rangle$ with $D$ its magnitude. All these terms are illustrated in Fig.~\ref{fig:model}. In this paper, we only consider the case that the vector $\boldsymbol{D}_{ij}$ is perpendicular to the lattice plane, namely only $D^{z}_{ij}$ is finite, so that total $S_z$ is conserved. We set $J_1 = 1$ as the energy unit for convenience. 

The model (\ref{eq:hamiltonian}) will be investigated by the variational Monte Carlo (VMC) method.  First,  we use the fermionic doublet representation to rewrite
the spin interactions as following,
\begin{equation}
	\begin{aligned}
		&\boldsymbol{S}_i \cdot \boldsymbol{S}_j = -\frac{1}{4}(T_{ij}T_{ij}^\dagger + P_{ij}P_{ij}^\dagger) + \mathrm{const},\\
		&\boldsymbol{S}_i \times \boldsymbol{S}_j = -\frac{i}{4}(T_{ij}\boldsymbol{T}_{ij}^\dagger + P_{ij}\boldsymbol{P}_{ij}^\dagger - \mathrm{h.c.}) + \mathrm{const},
	\end{aligned}
	\label{eq:fermionic_representation}
\end{equation}
where $T_{ij} = \psi^\dagger_i \psi_j$ ($P_{ij} = \psi^\dagger_i \bar{\psi}_j$) is the singlet hopping (pairing) term, while $\boldsymbol{T}_{ij} = \psi^\dagger_i \boldsymbol{\sigma} \psi_j$ ($\boldsymbol{P}_{ij} = \psi^\dagger_i \boldsymbol{\sigma} \bar{\psi}_j$) the triplet hopping (pairing) term.  The fermionic doublet field and its particle-hole partner are given by $\psi = (c_\uparrow, c_\downarrow){^\top}$ and  $\bar{\psi} = (c^\dagger_\downarrow, -c^\dagger_\uparrow){^\top}$, respectvely.
Considering the SU(2) gauge structure of this fermionic representation~\cite{Affleck-1988}, it is necessary to implement Lagrangian multipliers $\boldsymbol{\lambda}$ to enforce the generators of the SU(2) gauge group $\boldsymbol{\Lambda}_i = 0$ to return the subspace of real physical states. Their expressions with fermionic doublet representation are as following,
\begin{equation}
	\begin{aligned}
		&\Lambda_i^x = -\frac{1}{4}(\psi_i^\dagger \bar{\psi}_i + \bar{\psi}_i^\dagger \psi_i),\\
		&\Lambda_i^y = -\frac{i}{4}(\psi_i^\dagger \bar{\psi}_i - \bar{\psi}_i^\dagger \psi_i),\\
		&\Lambda_i^z = \frac{1}{2}(1 - \psi_i^\dagger \psi_i).
	\end{aligned}
\end{equation}

Then, we use the fermionic parton approximation to decouple the spin interactions into noninteracting quadratic structure, and obtain the mean-filed Hamiltonian  (irrelevant constants are omitted),
\begin{equation}
	\begin{aligned}
		H_{\mathrm{mf}}^{all} =& \sum_{i,j}( t_{ij}^s \psi_i^\dagger \psi_j +
		\boldsymbol{t}^t_{ij} \cdot \psi_i^\dagger \boldsymbol{\sigma} \psi_j +
		\Delta_{ij}^s \psi_i^\dagger \bar{\psi}_j \\
		&+ \boldsymbol{\Delta}^t_{ij} \cdot \psi_i^\dagger \boldsymbol{\sigma} \bar{\psi}_j
		+ \mathrm{H.c.}) \\
		&+ \sum_{i} \boldsymbol{\lambda} \cdot \boldsymbol{\Lambda}_i
		- \boldsymbol{M}_i \cdot \psi^\dagger_i  \boldsymbol{\sigma} \psi_i / 2,
	\end{aligned}
	\label{eq:full-mf-hamiltonian}
\end{equation}
where $t_{ij}^s$, $\boldsymbol{t}^t_{ij} \parallel \boldsymbol{D}_{ij}$ are spinon hopping parameters and $\Delta_{ij}^s$, $\boldsymbol{\Delta}^t_{ij} \parallel \boldsymbol{D}_{ij}$ are pairing ones, and a background field $\boldsymbol{M}_i$ is applied to induce a static magnetic order. Therefore, all the variational parameters are $\alpha^{all} = (t_{ij}^s, \boldsymbol{t}^t_{ij}, \Delta_{ij}^s, \boldsymbol{\Delta}^t_{ij}, \boldsymbol{\lambda}, \boldsymbol{M}_i)$.

Obviously, there must be various different ansatzes from the mean-field Hamiltonian Eq. (\ref{eq:full-mf-hamiltonian}) with a plenty of variational parameters. We selectively consider various singlet and triplet hopping terms and several pairing terms combined with the projective symmetry group\cite{Wen-psg-2002, lu-2011-psg, Bieri-psg-tri-2016} (PSG). 
 
Actually, we have found that the spinon-pairing instability (the allowed first NN triplet pairing term with complex numbers) is vanishingly small ($|\boldsymbol{\Delta}_{\langle ij \rangle}^t| / t_{\langle ij \rangle}^s< 10^{-3}$) in our calculation. So, we will ignore all  pairing terms reasonably. Moreover, the $\boldsymbol{\lambda}$ term can also be ignored because of vanishing pairing terms in the actual VMC procedure. In this way, we can obtain the reduced mean-field Hamiltonian, which is written as, 

\begin{equation}
	\begin{aligned}
		H_{\mathrm{mf}} =& \sum_{i,j}( t_{ij}^s \psi_i^\dagger \psi_j + \mathrm{H.c.}) +
		\sum_{\langle ij \rangle} (\boldsymbol{t}^t_{ij} \cdot \psi_i^\dagger \boldsymbol{\sigma} \psi_j + \mathrm{H.c.})\\
		& - \sum_i \boldsymbol{M}_i \cdot \psi^\dagger_i  \boldsymbol{\sigma} \psi_i / 2,
	\end{aligned}
	\label{eq:mf-hamiltonian}
\end{equation}
where $t_{ij}^s$, $\boldsymbol{t}^t_{ij} \parallel \boldsymbol{D}_{ij}$ are the singlet and triplet spinon hopping parameters, respectively. The former includes the first and second NN hopping terms and the later only include the first NN one.

With the mean-field ground-state wave function $|\mathrm{GS(\alpha)}\rangle_{\mathrm{mf}}$,  we can obtain a trial wave function $|\Phi(\alpha)\rangle = P_G |\mathrm{GS(\alpha)}\rangle_{\mathrm{mf}}$, where $P_G$ is the Gutzwiller projection to guarantee the single occupancy condition, and $\alpha = (t_{ij}^s, \boldsymbol{t}^t_{ij}, \boldsymbol{M}_i$) are variational parameters. We emphasize that these parameters could be complex numbers in principle, so the whole parameter range is in fact large. In the variational process, we employ the stochastic reconfiguration scheme~\cite{SR-prb-2000} to optimize so many parameters. We have considered various Z$_2$ QSLs, U(1) QSLs, $\boldsymbol{Q} = 0$  and $cuboc1$ magnetic orders as initial trial states selectively (see Appendix \ref{ap:mf_ansats} for details). We adopt the torus geometry with $L_1 = L_2 = 12$ for main results, where $L_{1,2}$ are the lengths along the two Bravais kagome-lattice vectors $\boldsymbol{a}_{1} = (1, 0)$ and $\boldsymbol{a}_{2} = (-1/2, \sqrt{3}/2)$.

\section{\label{sec:result}result}

In this paper, we do not intend to obtain the comprehensive phase diagram of the spin model in the kagome lattice defined by model (\ref{eq:hamiltonian}). Instead, we will focus on the study of the possible states stabilized or induced by the additional DM interaction $D$ and the AFM interactions $J_d$ across the diagonals of the hexagons of the kagome lattice, in the presence of the first, second and usual third NN spin couplings. 

\begin{figure}[ht]
 \raggedright
\subfigure{
    \begin{minipage}[h]{0.4\linewidth}
    \includegraphics[width = \linewidth]{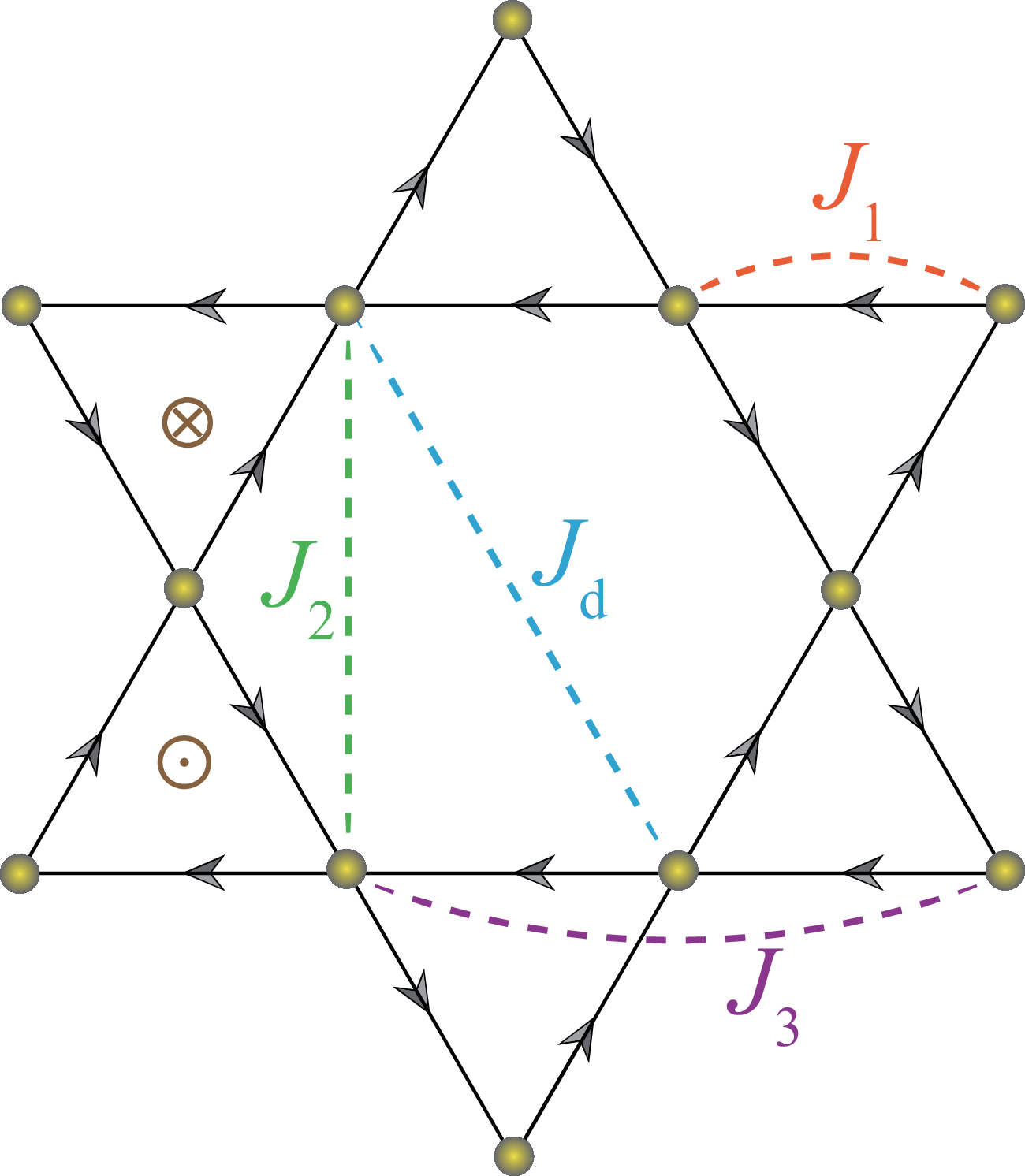}
    \put(-100, 100){(a)}
    \label{fig:model}
    \end{minipage}
    }%
\subfigure{
    \begin{minipage}[h]{0.52\linewidth}
    \includegraphics[width = \linewidth]{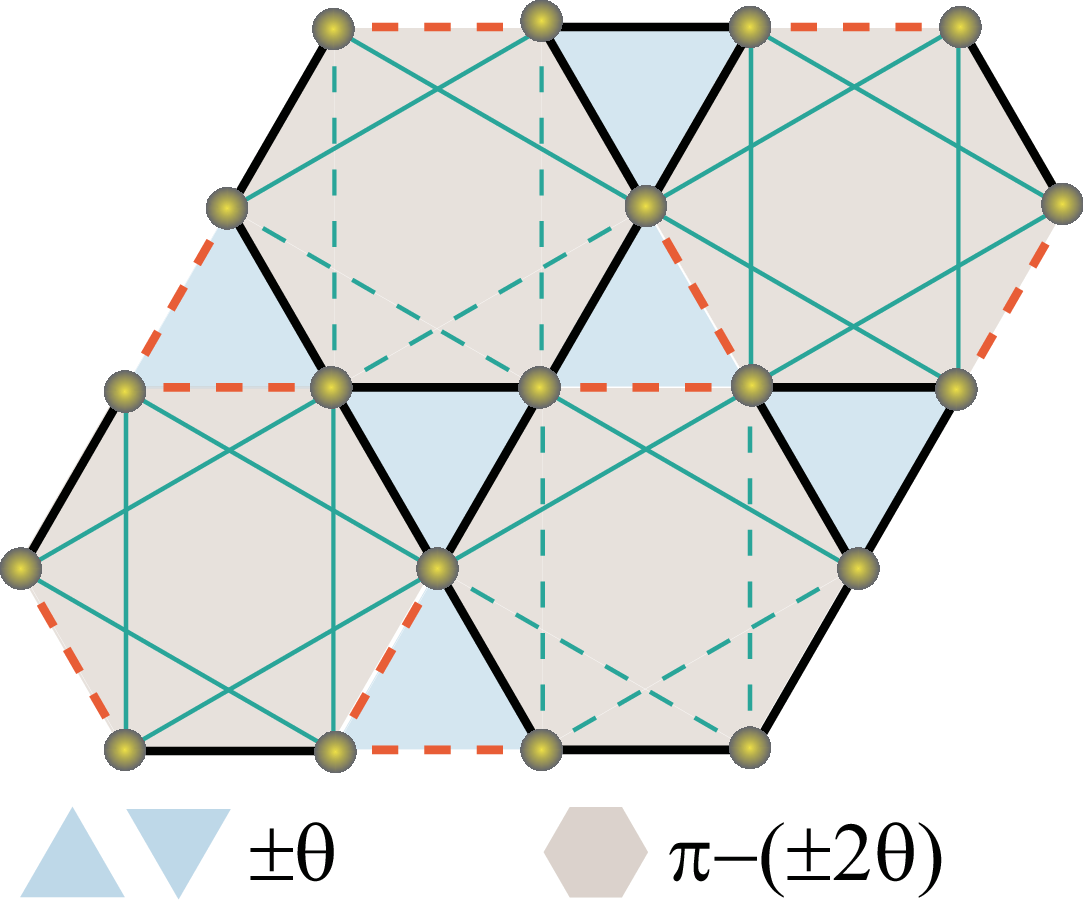}
    \put(-110, 100){(b)}
    \label{fig:ansatz_sqsh}
    \end{minipage}
    }
\subfigure{
    \begin{minipage}[h]{0.35\linewidth}
    \includegraphics[width = 1.0\linewidth]{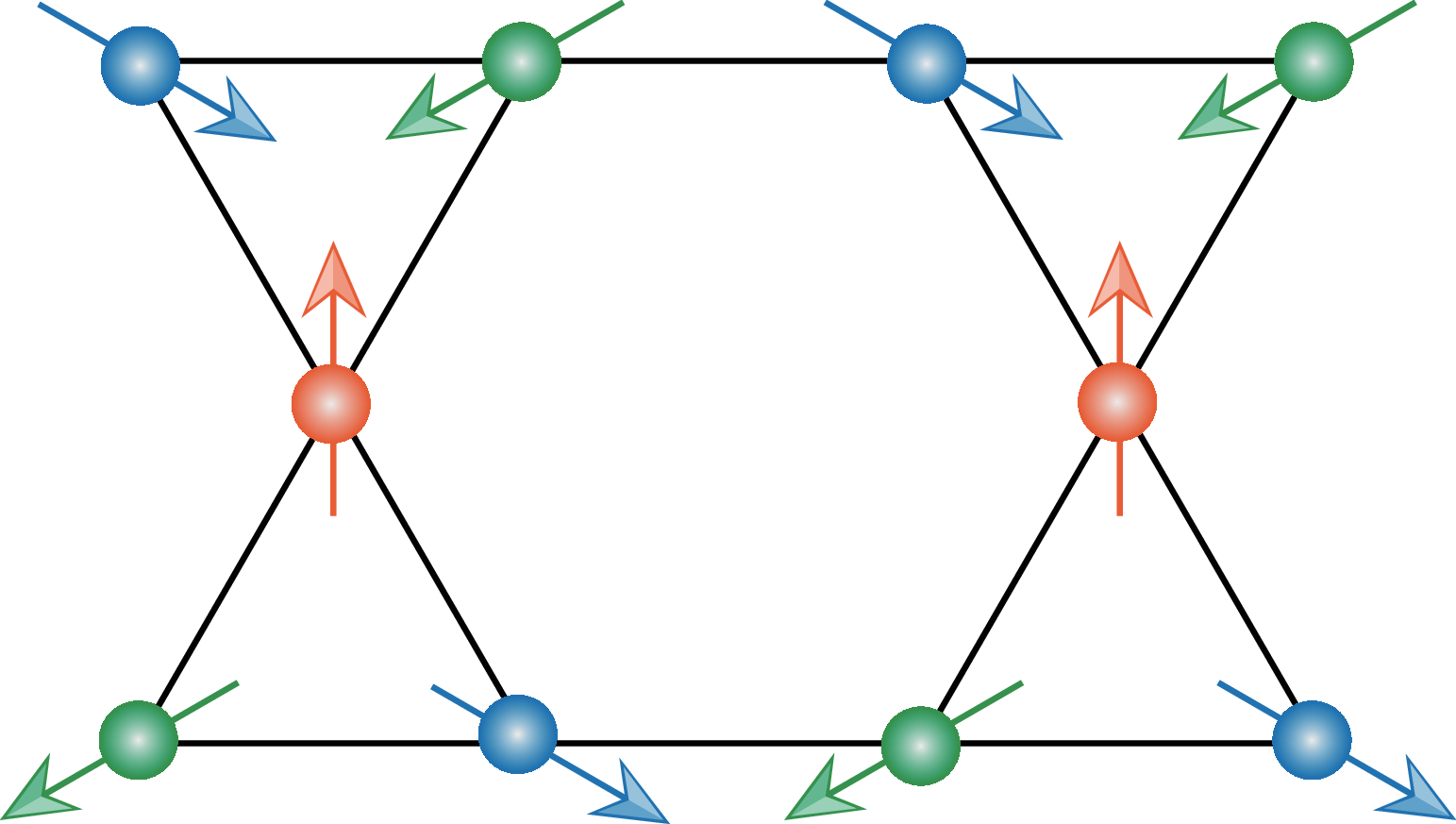}
    \put(-85, 55){(c)}
    \label{fig:ansatz_q0}
    \end{minipage}
    }%
\subfigure{
    \begin{minipage}[h]{0.62\linewidth}
    \includegraphics[width = 1.0\linewidth]{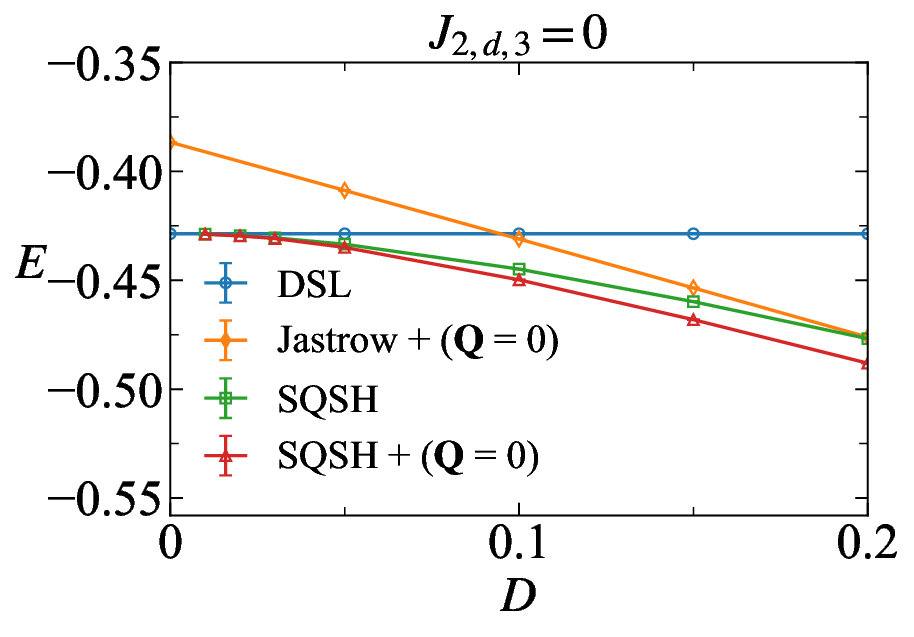}
    \put(-160, 90){(d)}
    \label{fig:en_vs_dm}
    \end{minipage}
    }
\caption{(a) Illustration of the spin interactions included in model (\ref{eq:hamiltonian}) on kagome lattice. The dashed lines with different colors indicate different Heisenberg terms. The arrows on the nearest-neighbor bonds indicate the direction of the DM interaction and the DM vector $\boldsymbol{D}_{ij}$ is oriented parallel (antiparallel) to $z$ axis in up (down) triangles indicated by $\bigodot$ ($\bigotimes$). (b) denotes the ansatz for the exotic SQSH state. The $+(-) \theta$ is the flux for spin-up (-down) spinon in each triangle, respectively, and the $\pi -(+) 2\theta$ is the same way in each hexagon. The marked red and blue dashed bonds denote that the hopping terms along these bonds in $H_\mathrm{mf}$ have the opposite sign compared with those unmarked ones so that the unit cell is doubled. (c) is the schematic diagram indicating the in-plane configuration of classical spins in the long-range magnetic order with $\boldsymbol{Q} = 0$,  three spins in all triangles form a $120^\circ$ distribution, which preserves the original translational symmetry. (d) illustrates the energy curves versus $D$ when $J_{2,d,3} = 0$ (see Appendix \ref{ap:mf_ansats} for the details of the four states). We note that all of the standard errors in this paper are considered as confidence intervals.}
\label{fig:model_ansatz}
\end{figure}

We start with the result in the presence of only the first NN $J_1$ AF couplings. Our VMC results suggest that a Dirac spin liquid is the most energetically favored state in this case, which is consistent with previous results\cite{Ran-dsl-2007, Iqbal-prb-j1j2-dsl, He-prx-dmrg-dsl}.  In this U(1) QSL, all triplet terms vanish and only the singlet hopping terms survive.  In the case of only the first NN hopping term is included, there are zero ($\pi$) fluxes through triangles (hexagons) in the kagome lattice, corresponding to the pattern for $\theta=0$ in Fig.~\ref{fig:ansatz_sqsh}.  
We also consider the Jastrow-type wave functions for the magnetic ordered state, which is widely used in VMC studies. This state is constructed based on the solution with only a finite $\boldsymbol{M}_i$ term in $H_\mathrm{mf}$ (see Appendix \ref{ap:mf_ansats}). It has the magnetic order $\boldsymbol{Q} = 0$ in the lattice (XY) plane as illustrated in Fig. \ref{fig:ansatz_q0} , and is called as Jastrow + $(\boldsymbol{Q}=0)$ state. In the case of only $J_1$ term, the energy of this magnetic order is obviously higher than that of the Dirac spin liquid, as shown in Fig. \ref{fig:en_vs_dm}.  

When turning on the DM interaction $D$,  we find that the energy of the Dirac spin liquid does not depend on $D$, while the energy of the Jastrow + $(\boldsymbol{Q}=0)$ state decreases nearly linearly with  $D$ (see Fig. \ref{fig:en_vs_dm}). However,  before the Jastrow + $(\boldsymbol{Q}=0)$ state surpasses energetically the Dirac spin liquid, a novel competing state emerges. This state has the same forms of the first and second NN singlet hopping terms as the Dirac spin liquid, but has a finite and pure imaginary first NN triplet hopping term.  The selection of this state is done via a comprehensive comparison with other possible states. We have considered the ansatzes with complex first and second NN singlet hopping terms,  but find that their imaginary parts are almost zero ($\mathrm{Im}(t_{\langle ij \rangle}^s) / t_{\langle ij \rangle}^s<10^{-2}$, $\mathrm{Im}(t_{\langle\langle ij \rangle\rangle}^s) / t_{\langle ij \rangle}^s<10^{-2}$) in our variational process.  We have also checked the existence of the first NN triplet pairing term (complex number), and it turns out that this term vanishes ($|\boldsymbol{\Delta}_{\langle ij \rangle}^t| / t_{\langle ij \rangle}^s< 10^{-3}$) in our calculation. In particular, we have considered two candidate states, which are regarded as the derivative states of the uniform resonating valence bond state (see Appendix \ref{ap:mf_ansats}).  One candidate carries the same fluxes in all triangles and the other carries the opposite fluxes in the up and down triangles. Combining the triplet hopping terms (complex number) and finite $\boldsymbol{M}_i$, we find that both these states are energetically unfavored.  

In this novel state, the complex triplet hopping term will lead to the consequence that free spinons at the VMC mean-field level will carry nonzero flux when hops along closed loops, as the direction of the DM vector $\boldsymbol{D}_{ij}$ considered in this paper is perpendicular to the lattice plane. Specifically, the spinon with up (down) spin carries a $+ (-) \theta$ flux in all triangles and $\pi - (+) 2 \theta$ flux in all hexagons. This shares the same physics as a quantum spin Hall state or topological insulator. So, we dub it as the spinon quantum spin Hall (SQSH) state and will discuss its properties in detail later. When $J_2=J_d=J_3=0$, this state emerges firstly at $D \sim 0.01$ in the sense that our numerical calculations can determine it. However, the mix of this state with the $\boldsymbol{Q} = 0$ magnetic order state, {\it i.e.} SQSH + ($\boldsymbol{Q} = 0$) state, is always lower in energy in the range $0.01 \alt D \alt 0.2$.  Moreover, this magnetic order state is energetically more favored than the Jastrow + ($\boldsymbol{Q} = 0$) state, as shown in Fig.~\ref{fig:en_vs_dm}.

The SQSH + ($\boldsymbol{Q} = 0$) state is induced and stabilized by the DM interaction. It is also analogous to the most possible ground state of the AFM $J_1$ Heisenberg model on the triangular lattice~\cite{Iqbal-j1j2-tri-2016, Zhao-Liu-j4-prl-2021, PhysRevB.108.245102}.  As the DM vector $\boldsymbol{D}_{ij}$ is considered to be along the perpendicular direction of the lattice plane, it confines the spins to lie in the lattice plane and effectively plays a role of the easy-plane anisotropy.  In the meantime, it is able to induce the local vector chirality $\boldsymbol{\chi}_\mathrm{v}$. These two properties make the DM interaction to favor the in-plane $\boldsymbol{Q} = 0$ magnetic order. In detail, this state requires a finite $\boldsymbol{M}_i$ and additionally the same hopping term as the SQSH state. However, due to the existence of the magnetic order,  the elementary excitation of this SQSH + ($\boldsymbol{Q} = 0$) state is the magnon which is the confined state of two spinons. Therefore, there are no free spinons existing in the sense of the VMC mean-field level and no SQSH effects.

\begin{figure}[ht]
	\centering
	\subfigure{
		\begin{minipage}[h]{0.98\linewidth}
			\includegraphics[width = 1.0\linewidth]{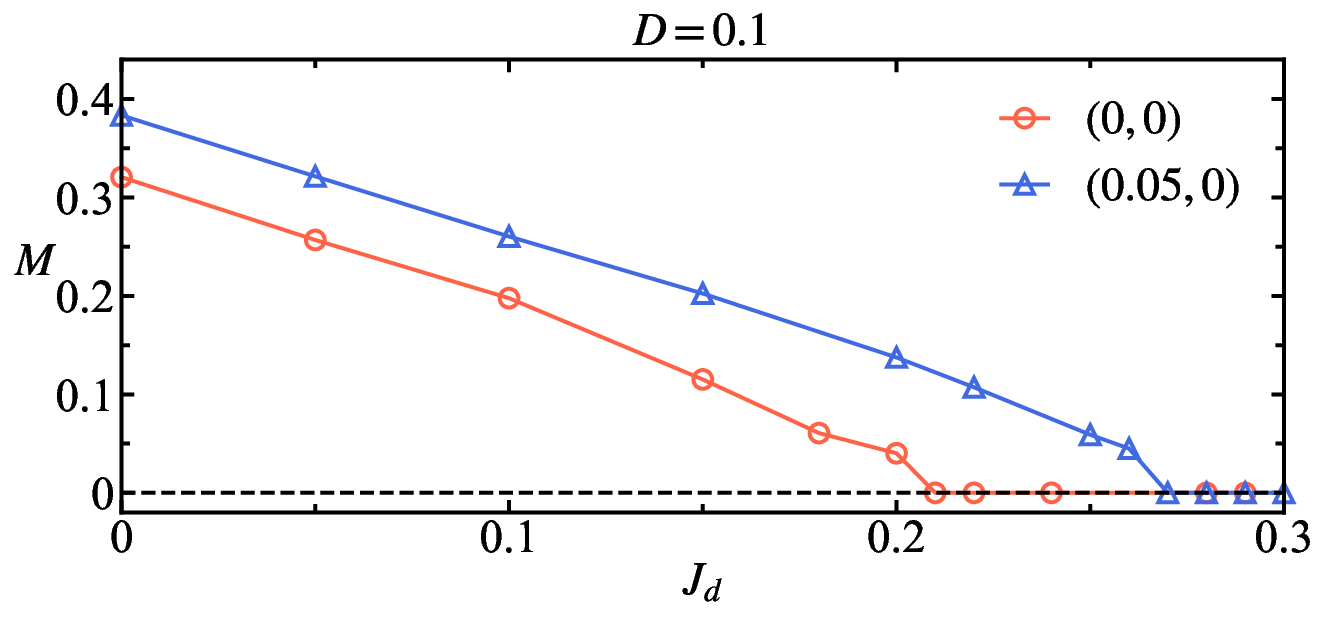}
			\put(-242, 100){(a)}
			\label{fig:m_vs_jd}
		\end{minipage}
	}
	\subfigure{
		\begin{minipage}[h]{0.47\linewidth}
			\includegraphics[width = 1.0\linewidth]{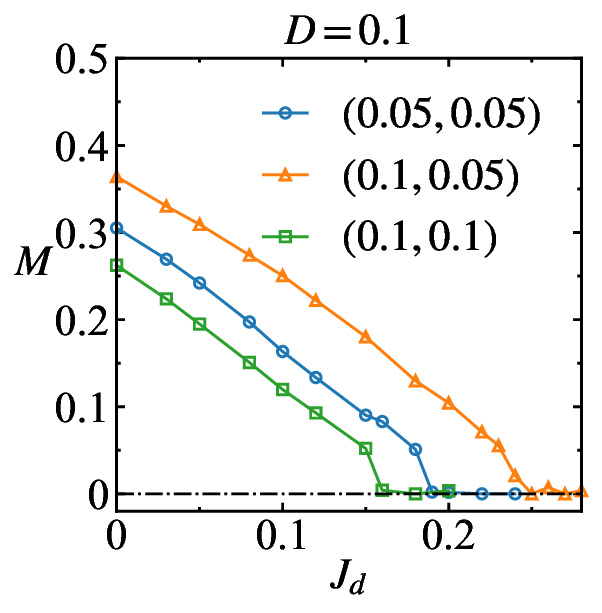}
			\put(-119, 105){(b)}
			\label{fig:m_vs_jd_dif1}
		\end{minipage}
	}%
	\subfigure{
		\begin{minipage}[h]{0.48\linewidth}
			\includegraphics[width = 1.0\linewidth]{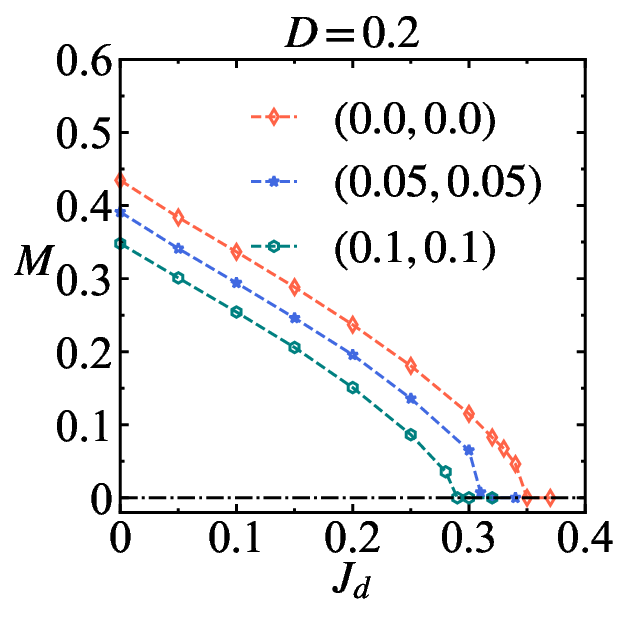}
			\put(-120, 104){(c)}
			\label{fig:m_vs_jd_dif2}
		\end{minipage}
	}
	\caption{Magnitude of magnetic moment ($M$) in the SQSH + ($\boldsymbol{Q} = 0$) state versus $J_d$ term for different $J_2,J_3$ interactions. (a) and (b) is obtained with the DM interaction $D=0.1$, (c) with $D=0.2$. We note that the legends of all these curves are labeled by $(J_2, J_3)$ and the lattice size we employ in (b) and (c) is $12 \times 6 \times 3$.}
	\label{fig:transition_q0-sqsh}
\end{figure}

It turns out that we need to melt the $\boldsymbol{Q} = 0$ magnetic order in order to realize SQSH state. We find that the AFM interaction $J_d$  across the diagonals of hexagons can effectively weaken the magnitude of the magnetic moment $M$. As can be seen in Fig.~\ref{fig:m_vs_jd},  $M$ decreases nearly linearly with $J_d$  and vanishes at $J_d \simeq 0.21$, when only the DM interaction besides the first NN term is considered, namely $D=0.1$ and $J_2=J_3=0$ [Fig.~\ref{fig:m_vs_jd})].
The introduction of the $J_2$ term enhances effectively the moment of the $\boldsymbol{Q} = 0$ magnetic order, so that a larger $J_d$ is needed to eliminate the magnetic order. As shown in Fig.~\ref{fig:m_vs_jd}, when $J_2$ starts from zero to 0.05, the critical $J^{c}_d$ term leading to $M \simeq 0$ increases to be 0.27. On the other hand, the effect of the $J_3$ term behaves conversely with $J_2$, but cooperatively with $J_d$ [Fig.~\ref{fig:m_vs_jd_dif1}]. 
For example, if the $J_2$ term is fixed to be $0.1$,  we can find that the moment $M$ with $J_3 = 0.1$ is significantly smaller than that with $J_3 = 0.05$ for the same $J_d$, and  $J^{c}_d$ drops to be near 0.16 for a fixed $J_3=J_2=0.1$.  To show more clearly the effect of the $J_3$ term, we present the results for the moment $M$  in the case of $D = 0.1$, $J_{2,d} = 0$ in Table~\ref{tab:q0_vs_j3}. To check the effects of the DM interaction, we show results obtained with $D=0.2$ in Fig.~\ref{fig:m_vs_jd_dif2}, one can see that $J^{c}_d$ increases noticeably compared to the case of $D=0.1$ for various $J_2$ and $J_3$. We note that those spin exchange parameters to melt the $\boldsymbol{Q} = 0$ magnetic order can be nearly one magnitude smaller than the dominant nearest-neighbor exchange interaction $J_1$, which are considered to be acceptable physically and realizable experimentally.

\begin{table}[ht]
	\centering
	\begin{tabular}{c|c|c|c|c|c}
		\hline \hline
		\ \  $J_3$  \ \ &  \ \ 0.05  \ \ & \ \ 0.08  \ \ &  \ \ 0.1  \ \ & \ \  0.12  \ \ &  \ \ 0.15 \ \ \\
		\hline
         \ \ $M$ \ \ & \ \ 0.2074 \ \ & \ \ 0.1239 \ \ & \ \ 0.0728 \ \ & \ \ 0.0279 \ \ & \ \ 0.0062 \ \ \\ 
		\hline \hline 
	\end{tabular}
	\caption{Magnetic moment $M$ in the SQSH + ($\boldsymbol{Q} = 0$) magnetic order state calculated with lattice size $12 \times 12 \times 3$ in the case of  $D = 0.1$, $J_{2,d} = 0$.}
	\label{tab:q0_vs_j3}
\end{table}

\begin{table}[ht]
	\centering
	\begin{tabular}{c|c|c|c|c}
		\hline \hline
		\ \ \ state \ \ \ & \ \ \ $D$ \ \ \ & \ \ \ $J_{2,3}$\ \ \ & \ \ \ $J_d$ \ \ \ & \ \ \ \ \ \ \ \ \ \ $E$ \ \ \ \ \ \ \ \ \ \ \\
		\hline
		\multirow{2}{*}{SQSH} 
		& 0.1 & 0.0 & 0.29 & -0.43371 \\
		& 0.1 & 0.0 & 0.3 & -0.43345 \\
		\hline
		\multirow{2}{*}{$cuboc1$} 
		& 0.1 & 0.0 & 0.29 & -0.43371 \\
		& 0.1 & 0.0 & 0.3 & -0.43379 \\
		\hline \hline 
	\end{tabular}
	\caption{Variational energy (per site) for the SQSH state and $cuboc1$ magnetically ordered state with lattice size $12 \times 12 \times 3$. When $D = 0.1$, $J_{2,3} = 0.0$ and $J_d \agt 0.3$, the $cuboc1$ order state with TR symmetry breaking will be dominant. All the errors of the energies are $\sim 10^{-5}$.}
	\label{tab:en_sqsh_vs_cuboc1_1}
\end{table}

\begin{table}[htbp]
	\centering
	\begin{tabular}{c|c|c|c|c|c}
		\hline \hline
		\ \ \ state \ \ \ & \ \ \ $D$ \ \ \ & \ \ \ $J_2$ \ \ \ & \ \ \ $J_d$ \ \ \ & \ \ \ $J_3$ \ \ \ & \ \ \ \ \ \ $E$ \ \ \ \ \ \ \\
		\hline
		\multirow{2}{*}{SQSH} 
		& 0.1 & 0.05 & 0.3 & 0 & -0.43444 \\
		& 0.1 & 0.05 & 0.32 & 0 & -0.43382 \\
		\hline
		\multirow{2}{*}{$cuboc1$} 
		& 0.1 & 0.05 & 0.3 & 0 & -0.43443 \\
		& 0.1 & 0.05 & 0.32 & 0 & -0.43387 \\
		\hline \hline 
	\end{tabular}
	\caption{Variational energy (per site) for the SQSH state and $cuboc1$ magnetically ordered state with lattice size $12 \times 12 \times 3$. When $D = 0.1$, $J_2 = 0.05$, $J_d \agt 0.32$ and $J_3 = 0.0$, the $cuboc1$ ordered state with TR symmetry breaking will be dominant. All the errors of the energies are $\sim 10^{-5}$.}
	\label{tab:en_sqsh_vs_cuboc1_2}
\end{table}

\begin{figure}[ht]
	\centering
	\subfigure{
		\begin{minipage}[ht]{0.85\linewidth}
			\includegraphics[width = 1.0\linewidth]{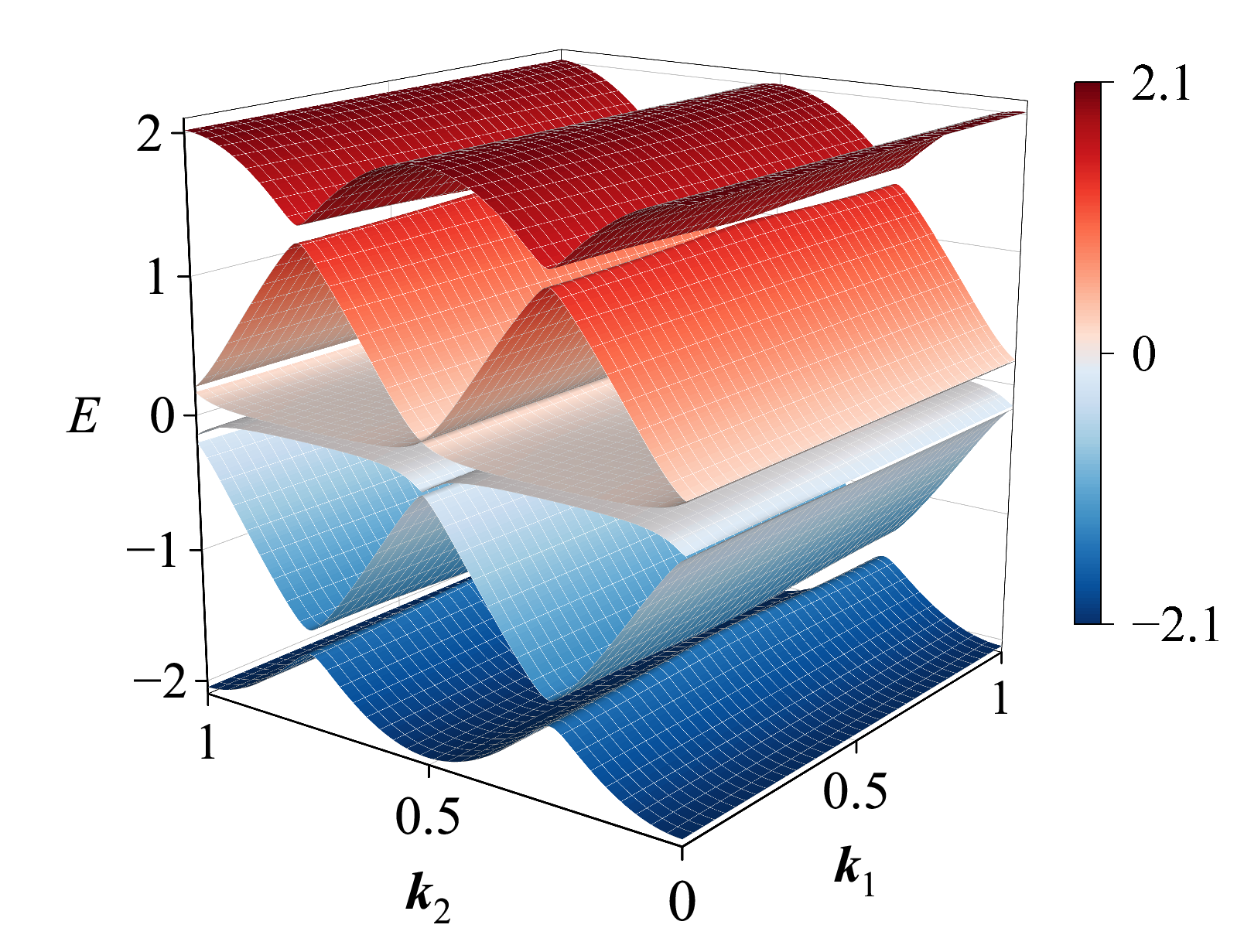}
			\put(-200, 138){(a)}
			\label{fig:ek_sqsh}
		\end{minipage}
	}
	\subfigure{
		\begin{minipage}[ht]{0.95\linewidth}
			\includegraphics[width = 1.0\linewidth]{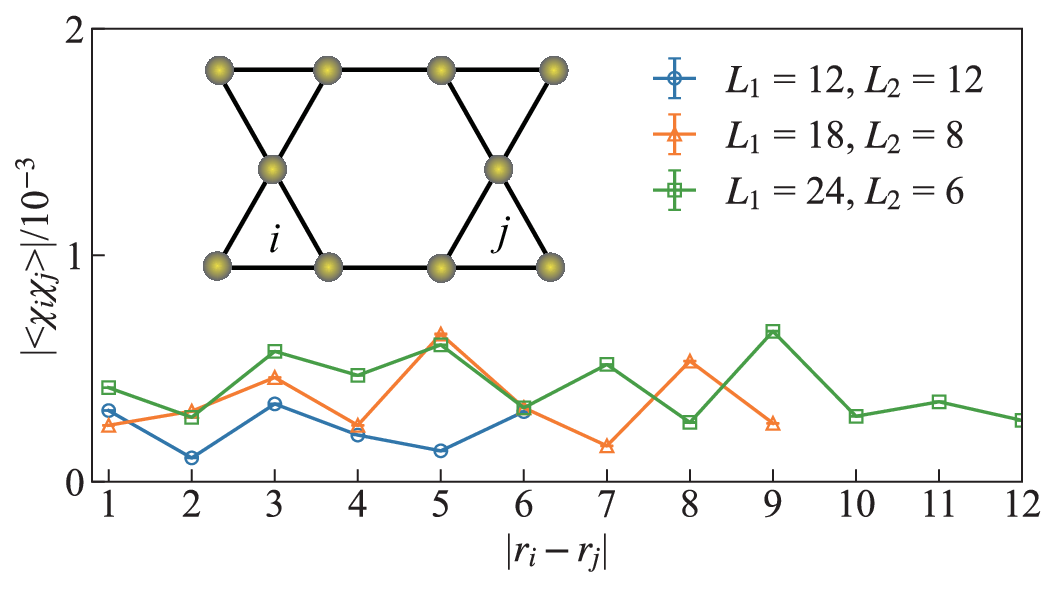}
			\put(-232, 124){(b)}
			\label{fig:chiral_cor}
		\end{minipage}
	}
	\caption{(a) Mean-field band structure of the SQSH state, where each band is doubly degenerate so that there is a significant energy gap because of the one spinon per site. (b)   The chirality-chirality correlation of the SQSH state $|\langle \chi_i \chi_j \rangle|$ ($i \neq j$) with respect to the distance along $\boldsymbol{a}_1$ for different lattice sizes. And the $|\boldsymbol{r}_i - \boldsymbol{r}_j|$ means the distance between two up-pointing triangles. We only need to consider the distance up to 6, 9 and 12 because of the period boundary condition for a system of $L_{1,2} = 12$, $L_1 = 18, L_2 = 8$ and $L_1 = 24, L_2 = 6$, respectively. These results are obtained in the system with $D=0.1$, $J_d = 0.21$ and $J_{2}=J_{3}=0$.}
	\label{fig:ek_and_chirality}
\end{figure}

\begin{figure}[htbp]
	\centering
	\includegraphics[width=\linewidth]{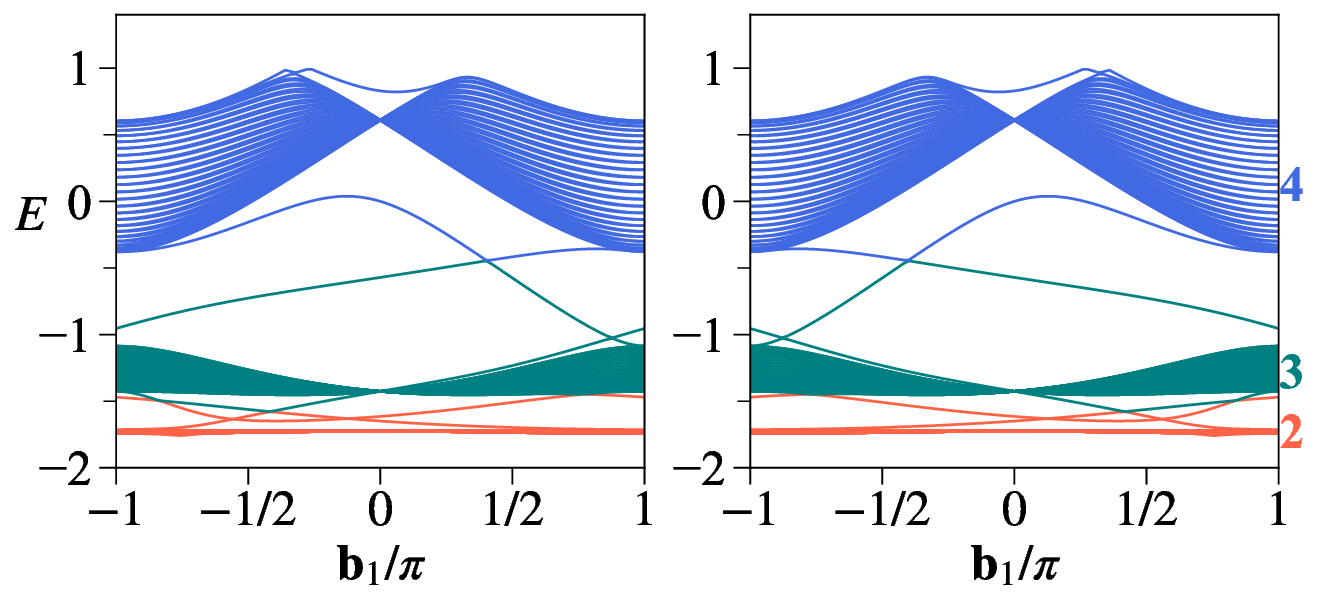}
	\caption{Left (right) panel is the spin-up (-down) energy dispersion of the SQSH state with the same optimal parameters as that of Fig.~\ref{fig:ek_sqsh}, respectively. In the calculation, the period (open) boundary condition is used in the direction of basis vectors $\boldsymbol{n}_1 = (2, 0)$ ($\boldsymbol{n}_2 = (-1/2, \sqrt{3}/2)$) defined in the doubled unit cell and the reciprocal vector $\mathbf{b}_1 = (\pi, \pi/\sqrt{3})$.}
	\label{fig:edg_sqsh}
\end{figure}

After the $\boldsymbol{Q} = 0$ magnetic order is melted, we further study the robustness of this disordered SQSH state against the $cuboc1$ order and present the results in Table \ref{tab:en_sqsh_vs_cuboc1_1} and \ref{tab:en_sqsh_vs_cuboc1_2} . It shows that the SQSH state can survive in a relatively broad range of $J_d$ and eventually gives way to the $cuboc1$ order, such as for $J_d \agt 0.3$. In addition, we find that the transition from the finite $M$ magnetic order state to the $M = 0$ spin liquid state with SQSH is a continuous phase transition. Now, let us discuss the detail properties of the pure SQSH state. As stated above, this state is a quantum spin liquid and inherits the same singlet hopping pattern as the Dirac quantum spin liquid [See Fig. \ref{fig:ansatz_sqsh}]. 
The key difference between them is that the SQSH state includes a pure imaginary $z$ component in the first NN triplet hoppings induced by the DM interaction. As a result,  the spin-up and -down spinons see the opposite flux ($\pm \theta$) in all triangles. Because of the absence of the coupling between spin-up and -down spinons, we can rewrite the mean-field Hamiltonian,
\begin{equation}
	H_{\mathrm{mf}} =
	\begin{pmatrix}
		h & 0\\
		0 & h^*
	\end{pmatrix},
	\label{eq:hmf_sqsh}
\end{equation}
where $h$ denotes the Hamiltonian for the spin-up spinons while $h^*$ that for the spin-down spinons. When we go into the $\boldsymbol{k}$ space, at an arbitrary $\boldsymbol{k}$, there is always a couple of degenerate states, which are conjugate to each other. So,  the spin-up and -down spinons have opposite Berry phase in any plaqutte of $\boldsymbol{k}$ space, and consequently the opposite Chern number. We have calculated the Chern number of the filled three bands for the spin-up (-down) spinons as shown in Fig. \ref{fig:ek_sqsh}, and find that $C_{\uparrow (\downarrow)} = + (-) 1$. We then calculate the energy dispersions of the spin-up (-down) spinons in the SQSH state with the period-open boundary condition, and the results are shown in Fig.~\ref{fig:edg_sqsh}. 

It shows clearly that two edge states emerge in the gap of the energy bands for each spin species spinon. In particular, the energy bands of the spin-up and -down spinons are antisymmetric with respect to momentum $k$, indicating that the spin-up state is just the time-reversal copy of the spin-down one. So, the spin-up and -down spinons move along opposite direction on the edge. We have also checked the chirality-chirality correlation in the SQSH state, the results are shown in Fig.~\ref{fig:chiral_cor}.  It shows $|\langle \chi_i \chi_j \rangle| \sim 0$ within numerical error, so the SQSH indeed has TR symmetry. It suggests that the two edge states of this topological SQSH are protected by TR symmetry.
Thus, we believe that the SQSH state is the spinon version of the quantum spin Hall state. 

\begin{figure}[htbp]
	\centering
	\includegraphics[width=0.95\linewidth]{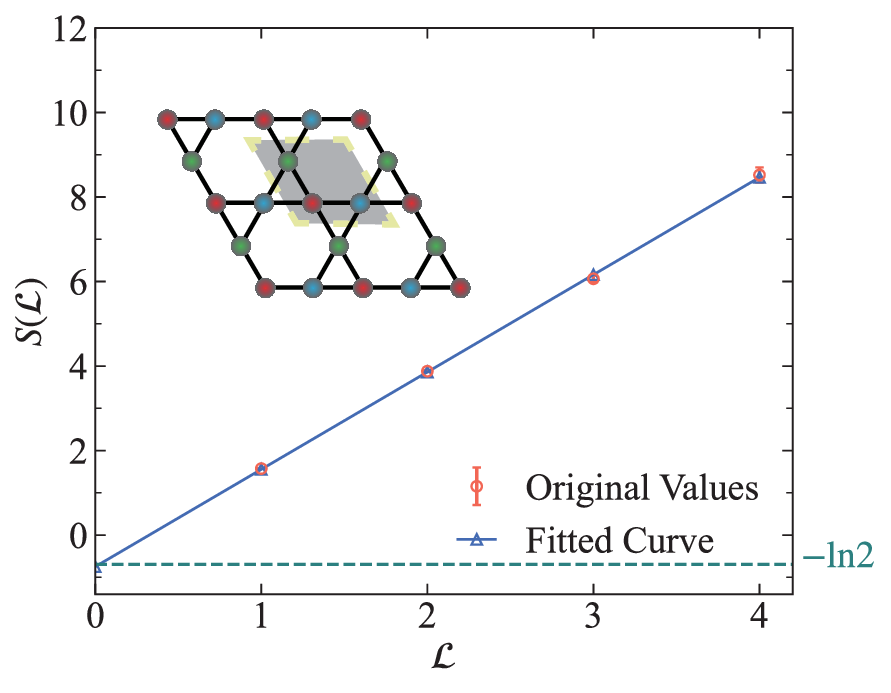}
	\caption{Entanglement entropy as a function of the area for the SQSH state with lattice size $12 \times 12 \times 3$. The corresponding optimal parameters are obtained when $D = 0.1$, $J_2 = 0.05$, $J_3 = 0$ and $J_d = 0.27$. The $x$-axis labeled as $\mathcal{L}$ means that the area is $\mathcal{L}^2$ times of the primitive cell, i.e., the shaded rhombus as shown in the insert, where different colored sites label the different sublattices of kagome lattice.}
	\label{fig:entropy}
\end{figure}

On the other hand, the SQSH state is essentially the strongly correlated state, which is different from the conventional quantum spin Hall state based on the single-particle picture. Its elementary excitations are fractionalized spinons out of the QSL ground state and could embrace more intrinsic topological properties. 
To further explore its property, we calculate the topological entanglement entropy and ground-state degeneracy. The TEE is obtained numerically 
by partitioning the system into two subsystems and calculating the second order Renyi entropy (see Appendix \ref{ap:gsd_tee} for details). Then, the Renyi entropy is expected to follow $S(\mathcal{L}) = \alpha \mathcal{L} - \gamma$, where $\alpha$ depends on the details of the state, $\mathcal{L}$ represents the boundary length of a contractible patch with codimension-1 boundary in the system and $\gamma$ is the universal TEE. In order to eliminate the area-law contribution $\alpha \mathcal{L}$, we calculate the entanglement entropy of plaquette $\mathrm{P}_1$ (the shaded region in the inset of Fig.~\ref{fig:entropy}) with different sizes, and the result is presented in Fig.~\ref{fig:entropy}.  It shows that $S(\mathcal{L})$ increases linearly with $\mathcal{L}$. Then, we apply a linear extrapolation to $\mathcal{L} \rightarrow 0$, and obtain the TEE for the SQSH state as $\gamma \approx  0.748$, which is very close to $\ln 2 = 0.693$. It shows that the SQSH state has intrinsic topological order. This is different from the quantum spin Hall state which is a symmetry-protected topological state. 

\begin{table}[htbp]
	\centering
	\begin{tabular}{c|c|c|c|c|c}
		\hline \hline
		\ \ state \ \ & \ \ $\varepsilon_1$ \ \ & \ \ $\varepsilon_2$ \ \ & \ \ $\varepsilon_3$ \ \ & \ \ $\varepsilon_4$ \ \ & \ \ $n_g$ \ \ \\
		\hline
		$\boldsymbol{Q} = 0$ & 3.997 & 1.3$\times 10^{-3}$ & 9.72$\times 10^{-4}$ & 9.27$\times 10^{-4}$ & 1\\
		\hline
		SQSH & 1.985 & 0.6774 & 0.673 & 0.6646 & 4\\
		\hline \hline 
	\end{tabular}
	\caption{Eigenvalues and GSDs of the $\boldsymbol{Q} = 0$ and SQSH states, calculated in a kagome lattice of 12$\times$12$\times$3. And $\varepsilon_i$ denotes the eigenvalue of  the overlap matrix while $n_g$ denotes GSD.}
	\label{tab:gsd}
\end{table}

The nontrivial topological properties of this state can be further characterized by its GSD. The calculation of GSD is carried out by constructing four states $|\phi_{\pm,\pm}\rangle_{\mathrm{mf}}$ with a Gutzwiller projection, where the + (-) subscript denotes the period (anti-period) boundary condition along the directions of two lattice basis vectors, respectively.  After diagonalizing the overlap matrix on the basis vector of these four states, we can obtain their eigenvalues, and the number of the significant finite eigenvalues is just GSD. In Table.~\ref{tab:gsd},  the results of four eigenvalues and the corresponding GSD are presented for the $\boldsymbol{Q} = 0$ magnetic order state and SQSH state. It shows that there is only one significant finite eigenvalue for the $\boldsymbol{Q} = 0$ state, so its GSD is $n_g = 1$. In the category of topological order~\cite{doi:10.1142/S0217979290000139, 10.1093/nsr/nwv077}, this magnetic order only holds a trivial (identity) topological excitation $\mathbb{I}$ with its quantum dimension $D_{q} = 1$ due to $n_g = 1$.  According to the relation $\gamma = \ln D_q$, we find its topological entanglement entropy is $\gamma = 0$. 
Besides, its Chern number is found to be zero. So, the $\boldsymbol{Q} = 0$ magnetic order state is a topological trivial phase without long-range entanglement with $\gamma = 0$. For the SQSH state, there are four finite eigenvalues and its GSD is $n_g = 4$.  From the above result of the TEE for the SQSH state $\gamma \approx  \ln 2$, we  obtain its quantum dimension $D_q=2$. For an Abelian topological phase, GSD = $D_q^2$. Therefore, we suggest that this SQSH state is an Abelian topological phase.

\section{\label{sec:conclusion}conclusions}
In summary, we have studied the interplay between the Dzyaloshinskii-Moriya interaction and the long-range AFM Heisenberg interactions in the kagome antiferromagnet by using the variational Monte Carlo method. We find that the Dzyaloshinskii-Moriya interaction alone favors the $\boldsymbol{Q} = 0$ long-range magnetic order, and an additional antiferromagnetic interaction across the diagonals of the hexagons of the kagome lattice can suppress and eliminate eventually this order phase. This topological phase transition leads to an exotic quantum spin state with fruitful topological properties. We elaborate that it is a topological gapped quantum spin liquid with the ground-state degeneracy $n_g = 4$ and the topological entanglement entropy $\gamma = \ln 2$. As the fractionalized excitations in a quantum spin liquid, the spinons constitute the two chiral edge states protected by the time-reversal symmetry. So, the spin-up and -down spinons move along opposite directions on a given edge, and it gives rise to the quantum spin Hall effect existing in a topological insulator.

We suggest that, by doping magnetic impurities into this spinon quantum spin Hall state to suppress one of the so-call helical states and retain the topological properties of the system as has been done for the celebrated quantum anomalous Hall effect~\cite{qah-sci-2013}, it is possible to detect the fractionalized spinons in this exotic quantum spin liquid.  

\begin{acknowledgments}
We would like to thank Q.-H. Wang, Z.-X. Liu, Z.-L. Gu, X.-M. Cui and J.-B. Liao for many helpful and valuable discussions. This work was supported by National Key Projects for Research and Development of China (Grant No. 2021YFA1400400) and the National Natural Science Foundation of China (No. 92165205).
\end{acknowledgments}

\appendix

\section{\label{ap:mf_ansats}MEAN-FIELD ANSATZES}

\begin{center} 
	{\bf 1. $\boldsymbol{Q}=0$ magnetic order}
\end{center}

Firstly, we consider a simple classical magnetic order $\boldsymbol{Q} = 0$ in the lattice (XY) plane. This state is constructed based on the solution with only a finite $\boldsymbol{M}_i$ term  in $H_\mathrm{mf}$, and is restricted to the $S_{tot}^{z}=0$ subspace~\cite{Iqbal-prb-j1j2-dsl} by the application of the projector $P_{S_{tot}^{z}=0}$.  Then, quantum fluctuations are included by the long-range Jastrow projector: 
\begin{equation}
    P_J^z = \mathrm{exp}\left(1/2 \sum_{ij} \mu_{ij} S_i^z S_j^z \right),
    \label{eq:jastrow}
\end{equation}
where $\mu_{ij}$ is the pseudopotential that only depends on the absolute distance $|\boldsymbol{R}_i - \boldsymbol{R}_j|$ of two sites. It decays exponentially with the distance so that we can just consider the first three without loss of generality. 
Finally, we get the state as $|\mathrm{\Psi_{\mathrm{mag}}}\rangle = P_{S_{tot}^{z}=0} P_J^z |\mathrm{GS}\rangle_{\mathrm{mf}}$, and call it the Jastrow + $(\boldsymbol{Q}=0)$ state.
\begin{center}
\bf 2. DSL
\end{center}

The Dirac quantum spin liquid (DSL) is one of the most competitive candidate ground states in the Heisenberg model with only $J_1$ term on kagome lattice. In this U(1) QSL, all triplet terms vanish and only singlet hopping terms survive. If we only consider the first NN hopping ones, there are zero ($\pi$) fluxes through triangles (hexagons), respectively. Based on PSG, the second singlet hopping terms can also exist. All bond patterns are shown in Fig.~\ref{fig:ansatz_sqsh} in the main text. We also emphasize that the third NN terms in this state are forbidden by PSG, so that we rationally abandon them. In fact, the second NN singlet pairing terms are allowed by PSG, namely, the so-called Z$_2[0, \pi]\beta$, a gapped QSL~\cite{lu-2011-psg}. But, it's not energy favorable in our model by our calculation. In detail, our numerical results suggest that the $J_1$, $J_d$ and $J_3$ interactions suppress this pairing term, which is consist with Ref.~\cite{Iqbal-prb-j1j2-dsl}. For this reason, we also throw away the second NN singlet pairing terms all the time.  

\begin{center}
	\bf 3. SQSH
\end{center}

The SQSH state is a novel state we found in this paper. In this state, the first and second NN singlet hopping terms have the same forms as those of the DSL discussed above, but the first NN triplet hopping term is finite and pure imaginary number. According to the direction of the DM interaction vector $\boldsymbol{D}_{ij}$ we chosen in this paper, this triplet hopping term will lead to that
the spin-up (-down) spinon sees $+ (-) \theta$ flux in all triangles and $\pi - (+) 2 \theta$ flux in all hexagons. Intrinsically, the spin-up and -down spinons have opposite Chern number $C_{\uparrow,\downarrow} = \pm 1$. 

As shown in Fig.~\ref{fig:en_vs_dm}, when $0.01 \alt D < 0.2$ and other interactions beyond $J_1$ term are absent, another magnetic order state as a magnetic instability of SQSH, we call it the SQSH + ($\boldsymbol{Q} = 0$) state, is energetically favored. For convenience, we stipulate the unique $\boldsymbol{Q} = 0$ magnetic ordered state mentioned in the main text is just the SQSH + ($\boldsymbol{Q} = 0$) state, because the Jastrow + ($\boldsymbol{Q} = 0$) state is not energy favorable. Henceforth, we also call it the $\boldsymbol{Q} = 0$ state.

\begin{center}
	\bf 4. CSL
\end{center}

We have considered the chiral quantum spin liquid (CSL), in which the first and second NN singlet hopping terms are complex numbers as discussed in Ref.~\cite{hu-j12d_ch-prb-2015-csl}. But, we find that their imaginary parts are almost zero ($\mathrm{Im}(t_{\langle ij \rangle}^s) / t_{\langle ij \rangle}^s<10^{-2}$, $\mathrm{Im}(t_{\langle\langle ij \rangle\rangle}^s) / t_{\langle ij \rangle}^s<10^{-2}$) in our variational process. At least, in the range of interactions we considered in this work, this result excludes the chiral QSL stabilized by strong enough $J_d$~\cite{hu-j12d_ch-prb-2015-csl, j12d_jch-prb-2015-csl}. 

\begin{center}
	\bf 5.  $cuboc1$ magnetic order
\end{center}

\begin{figure}[ht]
	\centering
	\includegraphics[width=0.8\linewidth]{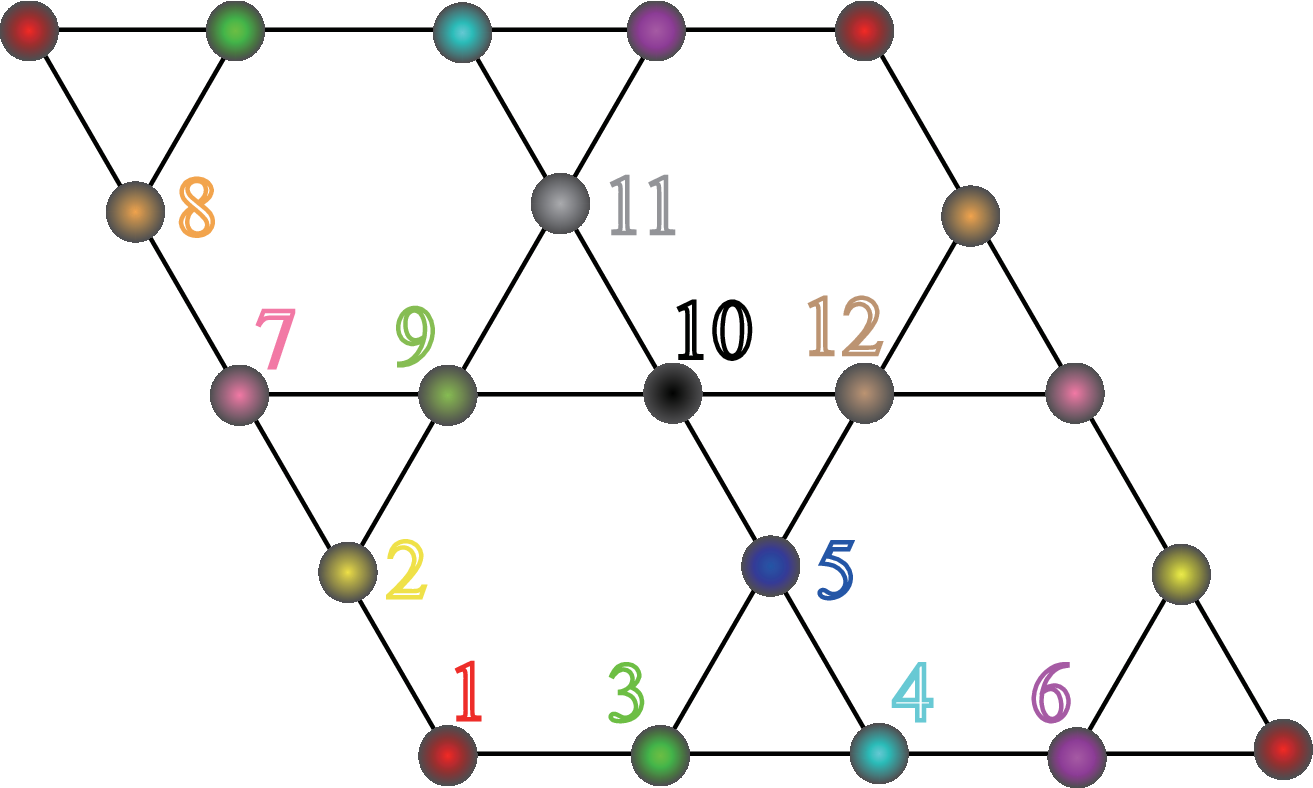}
	\caption{12 sublattices of the magnetic unit cell for classical $cuboc1$ order. $\boldsymbol{S}_1 = (1, 0, 0)$, $\boldsymbol{S}_2 = (-\frac{1}{2}, \frac{\sqrt{3}}{2}, 0)$, $\boldsymbol{S}_3 = (-\frac{1}{2}, \frac{\sqrt{3}}{6}, \frac{\sqrt{6}}{3})$, $\boldsymbol{S}_4 = (0, \frac{\sqrt{3}}{3}, -\frac{\sqrt{6}}{3})$, $\boldsymbol{S}_5 = (\frac{1}{2}, -\frac{\sqrt{3}}{2}, 0)$, $\boldsymbol{S}_6 = (-\frac{1}{2}, -\frac{\sqrt{3}}{2}, 0)$, $\boldsymbol{S}_7 = (1, -\frac{\sqrt{3}}{3}, \frac{\sqrt{6}}{3})$, $\boldsymbol{S}_8 = (-\frac{1}{2}, -\frac{\sqrt{3}}{6}, -\frac{\sqrt{6}}{3})$, $\boldsymbol{S}_9 = (\frac{1}{2}, -\frac{\sqrt{3}}{6}, -\frac{\sqrt{6}}{3})$, $\boldsymbol{S}_{10} = (-1, 0, 0)$, $\boldsymbol{S}_{11} = (\frac{1}{2}, \frac{\sqrt{3}}{6}, \frac{\sqrt{6}}{3})$ and $\boldsymbol{S}_{12} = (\frac{1}{2}, \frac{\sqrt{3}}{2}, 0)$.}
	\label{fig:cuboc1}
\end{figure}

We also study the so-called $cuboc1$ magnetic order as the instability of the TR symmetry breaking chiral spin liquid, i.e., CSL + $cuboc1$~\cite{j12d-prb-phase-diagram-2015}. Without loss of generality, we additionally include the first NN triple hopping term $t_{<ij>,z}^t$ into this ansatz. The detail of the classical $cuboc1$ order is shown in Fig.~\ref{fig:cuboc1}. For the sake of simplicity, we call it $cuboc1$ state. We indeed find that the variational energy of this magnetic order is lower than that of the SQSH state, when the $J_d$ term is relatively large, as listed in Table~\ref{tab:en_sqsh_vs_cuboc1_1} and \ref{tab:en_sqsh_vs_cuboc1_2}. Before the phase transition, we find that the $cuboc1$ order almost reduces to SQSH state by VMC calculation because the optimal parameters $t_{<ij>}^s$ and $t_{<<ij>>}^s$ are real number, the optimal parameter $t_{<ij>,z}^t$ is pure imaginary and the magnetic moment $M$ is almost vanishing. As a result, the energies of both $cuboc1$ order and SQSH state look like degeneracy within numerical error. Therefore, we believe this exotic SQSH state can survive in a relatively broad range of $J_d$.

\begin{center}
	\bf 6. uRVB state
\end{center}

In addition, we have considered another two candidate states. Both of them are regarded as the derivative states of uRVB states. Without loss in generality, we allow the first NN singlet $t_{ij}^s$ terms are complex number so that there are gauge fluxes though triangular and hexagonal plaquette. And we consider two flux patterns, one where fluxes through all triangles are the same and another one where fluxes through up and down triangles are the opposite. Combining the triplet hopping $t_{ij,z}^t$ terms (complex number) and finite $\boldsymbol{M}_i$ of the $\boldsymbol{Q} = 0$ magnetic order, we find that the trial energies of both states are much higher than aforementioned states. So, we discard them and do not draw up their energy curves in Fig.~\ref{fig:en_vs_dm} in main text.

\section{\label{ap:chern_number} CHERN NUMBER}
Nonzero Chern number is one of the fundamental topological numbers to characterize topological phase of matter. Here we won't go into details about the concepts of Berry connection, Berry phase and Chern number with formal analytical expression. We just introduce the numerical calculation of Chern number in the filled bands~\cite{chern_num_2005}.

Firstly, we obtain the corresponding mean-field Hamiltonian with the optimized variational parameters by VMC and then transform it into $\boldsymbol{k}$-space, $H(\boldsymbol{k})$. And we  emphasize that the $H(\boldsymbol{k})$ is periodic along the directions of reciprocal basis vectors $\boldsymbol{b}_{1,2}$, $H(\boldsymbol{k}) = H(\boldsymbol{k} + n_1 \boldsymbol{b}_1 + n_2 \boldsymbol{b}_2)$, where $n_{1,2}$ are any integer numbers. In other word, $H(\boldsymbol{k})$ is in Bloch form. Therefore, this Fourier transformation must be handled with care. To be specific, it usually needs a gauge transformation, $c_{\boldsymbol{k}} \longrightarrow c_{\boldsymbol{k}} e^{i\boldsymbol{k} \cdot \boldsymbol{\delta}}$. In general, the $\boldsymbol{\delta}$ is different for different $c_{\boldsymbol{k}}$ and not unique.

Then, for a lattice with finite size, the Brillouin zone is filled with discrete $\boldsymbol{k}$ points. And we define intervals of $\boldsymbol{k}$ points in two directions of reciprocal basis vectors,
\begin{equation}
    \boldsymbol{u}_i = \frac{l_{i} \boldsymbol{b}_{i}}{2N_{i}\pi}, \quad (i = 1,2; \quad N_i / l_i \in \boldsymbol{\mathrm{N}}^* ).
    \label{eq:k_interval}
\end{equation}
In our calculation, we take $l_i = 1$ to guarantee the highest numerical precision. We also note larger intervals are also allowed as long as the result is convergent. And then, we require that the eigenstate $|n(\boldsymbol{k}) \rangle$ of $H(\boldsymbol{k})$ is also periodic in the Brillouin zone to eliminate the effect of any U(1) gauge of eigenstate. Now we can define the U(1) quantity for a certain $\boldsymbol{k}$ as following,
\begin{equation}
    \eta(\boldsymbol{k})_{\boldsymbol{u}_i} \equiv \frac{\langle n(\boldsymbol{k})| n(\boldsymbol{k} + \boldsymbol{u}_i) \rangle}{|\langle n(\boldsymbol{k})| n(\boldsymbol{k} + \boldsymbol{u}_i) \rangle|}.
    \label{eq:e_itheta}
\end{equation}
$\eta(\boldsymbol{k})_{\boldsymbol{u}_i}$ is well defined as long as the denominator of Eq. \ref{eq:e_itheta} is nonzero. And then, we can define another variable about phase in a loop with $\eta(\boldsymbol{k})_{\boldsymbol{u}_i}$,
\begin{equation}
    \begin{aligned}
        &\theta(\boldsymbol{k}) = \frac{1}{i}\ln\left(\eta(\boldsymbol{k})_{\boldsymbol{u}_1} \eta(\boldsymbol{k} + \boldsymbol{u}_1)_{\boldsymbol{u}_2} {\eta(\boldsymbol{k} + \boldsymbol{u}_2)^\dagger_{\boldsymbol{u}_1}} {\eta(\boldsymbol{k})^\dagger_{\boldsymbol{u}_2}}\right),\\
        &-\pi < \theta(\boldsymbol{k}) \leq \pi.    
    \end{aligned}
    \label{eq:berry_phase}
\end{equation}
Finally, the Chern number of the $n$th filled band is obtained by,
\begin{equation}
    C_n \equiv \frac{1}{2\pi} \sum_{\boldsymbol{k} \in \mathrm{BZ}} \theta(\boldsymbol{k}).
    \label{eq:chern_number}
\end{equation}

\section{\label{ap:gsd_tee} GROUND-STATE DEGENERACY AND TOPOLOGICAL ENTANGLEMENT ENTROPY}
The low-energy gauge fluctuations of gapped QSLs (topological orders) are characterized by ground-state degeneracy (GSD). When one compacts the lattice to a torus, in the thermodynamics limit, there is no energy cost when a Z$_\mathrm{2}$ $\pi$ flux is inserted in any hole of the torus. In the mean-field theory, this procedure is equivalent to changing the boundary condition of $H_\mathrm{mf}$ from period to anti-period. For a two-dimensional system, in general, we can always construct four states $|\phi_{\pm,\pm}\rangle_{\mathrm{mf}}$, where the + (-) subscript denotes the periodic (anti-periodic) boundary condition along the directions of two lattice basis vectors, respectively. Then, we enforce a Gutzwiller projection to these four ground states of $H_\mathrm{mf}$ to recover physical Hilbert space. Thus, we rewrite the symbols of the four states for convenience,
\begin{equation}
    \begin{aligned}
        &|1\rangle = P_G |\phi_{+,+}\rangle_{\mathrm{mf}}, \quad
        |2\rangle = P_G |\phi_{+,-}\rangle_{\mathrm{mf}},\\
        &|3\rangle = P_G |\phi_{-,+}\rangle_{\mathrm{mf}}, \quad
        |4\rangle = P_G |\phi_{-,-}\rangle_{\mathrm{mf}}.
    \end{aligned}
    \label{eq:four_state_gsd}
\end{equation}
We can calculate the 4 by 4 overlap matrix $\mathcal{O}$ based on these four states. In detail, the matrix element $\mathcal{O}_{ij} = \langle i | j \rangle / \sqrt{\langle i | i \rangle \langle j | j \rangle}$, where $i,j = 1, 2, 3, 4$. After diagonalizing this overlap matrix, we can obtain its eigenvalues. The number of the significantly finite eigenvalues is just GSD. We calculate the overlap matrices of the $\boldsymbol{Q} = 0$ magnetically ordered state and SQSH states as following,
\begin{equation}
\begin{aligned}
    &\mathcal{O}_{\boldsymbol{Q} = 0} \simeq 
    \begin{pmatrix}
    1 & e^{-i0.29} & e^{-i0.79} & e^{-i0.76}\\
    e^{i0.29} & 1 & e^{-i0.5} & e^{-i0.46}\\
    e^{i0.79} & e^{i0.5} & 1 & e^{i0.03}\\
    e^{i0.76} & e^{i0.46} & e^{-i0.03} & 1
    \end{pmatrix},\\
    &\mathcal{O}_{\mathrm{SQSH}} \simeq \\
    &\begin{pmatrix}
    1 & 0.32e^{i2.64} & 0.33e^{-i2.13} & 0.33e^{i3.05}\\
    0.32e^{-i2.64} & 1 & 0.32e^{i1.51} & 0.33e^{i0.42}\\
    0.33e^{i2.13} & 0.32e^{-i1.51} & 1 & 0.33e^{-i1.1}\\
    0.33e^{-i3.05} & 0.33e^{-i0.42} & 0.33e^{i1.1} & 1
    \end{pmatrix}.
\end{aligned}
\label{eq:overlap_q0_sqsh}
\end{equation}
And now, we can obtain their eigenvalues and GSDs, as shown in Table~\ref{tab:gsd}.

\begin{figure}[htbp]
     \centering
\subfigure{
    \begin{minipage}[h]{0.3\linewidth}
    \includegraphics[width = 1.0\linewidth]{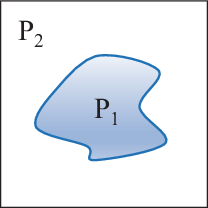}
    \put(-40, -10){(a)}
    \label{fig:bipartition}
    \end{minipage}
    }
\subfigure{
    \begin{minipage}[h]{0.5\linewidth}
    \includegraphics[width = 1.0\linewidth]{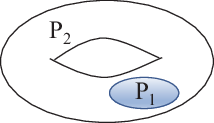}
    \put(-68, -12){(b)}
    \label{fig:torus}
    \end{minipage}
    }
    \caption{(a) shows a schematic diagrm of bipartition on an arbitrary system. (b) indicates the system is compacted to a torus. And this bipartition is trivial so that the boundaries are contractile.}
    \label{fig:bipartition_and_torus}
\end{figure}

Another important quantity to characterize topological order is topological entanglement entropy (TEE)~\cite{tee-levin-wen-2006-prl, tee-kitaev-2006-prl}. Firstly, we divide a system into two parts, as shown in Fig. \ref{fig:bipartition}.
And then, the von Neumann entanglement entropy of P$_\mathrm{1}$ for a system can be represented as follows,
\begin{equation}
    S({\mathrm{P_1}}) = \alpha \mathcal{L} - \gamma,
    \label{eq:entropy_p1}
\end{equation}
where the coefficient $\alpha$ is not universal and depends on the details of the state, $\mathcal{L}$ is the codimensional-1 boundary of P$_\mathrm{1}$, and $\gamma$ is just the universal TEE. As we know, $S({\mathrm{P_1}}) \geq 0$ for an arbitrary gapped system~\cite{Savary-rev-qsl-2017}. For those systems with $\gamma = 0$, it is possible to deform the ground state to obtain $\alpha = 0$, namely one can remove the leading area-low contribution ($\alpha \mathcal{L} = 0$). Thus, the state with $\alpha = 0$ is the pure direct product one without long-range entanglement, i.e. non-topological gapped phase. On the contrary, the deformation could never occur for the state with $\gamma > 0$, which is the case for a topological gapped phase. Therefore, the nonzero $\gamma$ is a direct sign of long-range entanglement.

Numerical calculation of TEE from the von Neumann entropy is difficult, so we focus on the Renyi entropy here. The Renyi entropy for the gapped state with bipartition is defined as~\cite{Zhang-tee-2011},
\begin{equation}
    S_n = \frac{1}{1-n} \ln \left[ \mathrm{Tr}\left(\rho_1^n \right) \right],
\end{equation}
where $\rho_1$ is the reduced density matrix obtained by tracing out the subsystem P$_\mathrm{2}$, $\rho_1 = \mathrm{Tr}_2 |\Psi \rangle \langle \Psi|$, where $|\Psi \rangle$ is a normalized wave function of the system. In this paper, we just focus on the Renyi entropy with index $n = 2$, $S_2 = - \ln \left[ \mathrm{Tr}\left(\rho_1^2 \right) \right]$. We define a swap operator $X$\cite{renyi-swaps-prl-2010} with the purpose as follow,
\begin{equation}
    X|\alpha_1 \rangle \otimes |\alpha_2 \rangle = |\beta_1 \rangle \otimes |\beta_2 \rangle,
    \label{eq:swap_operator}
\end{equation}
where $|\alpha_1 \rangle = |a \rangle |b \rangle$ and $|\alpha_2 \rangle = |m \rangle |n \rangle$ are two configurations, the $|a \rangle$ and $|m \rangle$ are in P$_\mathrm{1}$ while $|b \rangle$ and $|n \rangle$ are in P$_\mathrm{2}$, and then $|\beta_1 \rangle = |m \rangle |b \rangle$ and $|\beta_2 \rangle = |a \rangle |n \rangle$. 

Finally, we can rewrite $S_2$ in terms of the expectation of $X$ with respect to the wave function $|\Psi \rangle \otimes |\Psi \rangle$, $S_2 = -\ln \langle X \rangle$. Empirically, $\langle X \rangle$ is predicted to be a complex number in actual calculation if  $|\Psi \rangle$ is complex. So we can divide this expectation into two parts, $\langle X \rangle = \langle X_{mod} \rangle \langle X_{phase} \rangle$, which can be individually calculated by Monte Carlo (MC) method as shown in Eq.~\ref{eq:swap_expectation_mc}. It is worth mentioning that $\tilde{\rho}_{\alpha_1, \alpha_2}$ is a joint probability distribution. Besides, we note that for large size $\mathcal{L}$, the computational cost is relatively high because we have to taken more samples in the MC process to reduce numerical error. Therefore, we calculate the $S(\mathcal{L})$ with $\mathcal{L} = 1$ up to $4$ to eliminate the area contribution and obtain the TEE.
\begin{equation}
    \begin{aligned}
        &\langle X_{mod} \rangle = \sum_{\alpha_1, \alpha_2} \rho_{\alpha_1} \rho_{\alpha_2} \left| f(\alpha_1, \alpha_2) \right|,\\  
        &\langle X_{phase} \rangle = \sum_{\alpha_1, \alpha_2} \tilde{\rho}_{\alpha_1, \alpha_2} e^{i\theta(\alpha_1, \alpha_2)},\\
        &\rho_{\alpha_i} = \frac{\left|\langle \alpha_i | \Psi \rangle \right|^2}{\langle \Psi | \Psi \rangle},
        f(\alpha_1, \alpha_2) = \frac{\langle \beta_1 | \Psi \rangle \langle \beta_2 | \Psi \rangle}{\langle \alpha_1 | \Psi \rangle \langle \alpha_2 | \Psi \rangle},\\
        &\tilde{\rho}_{\alpha_1, \alpha_2} = \frac{\left| \langle \alpha_1 | \Psi \rangle \langle \alpha_2 | \Psi \rangle \right|^2 \left| f(\alpha_1, \alpha_2) \right|}{\sum_{\alpha_1, \alpha_2} \left| \langle \alpha_1 | \Psi \rangle \langle \alpha_2 | \Psi \rangle \right|^2 \left| f(\alpha_1, \alpha_2) \right|},\\
        &e^{i\theta(\alpha_1, \alpha_2)} = \frac{f(\alpha_1, \alpha_2)}{\left| f(\alpha_1, \alpha_2) \right|}.
    \end{aligned}
    \label{eq:swap_expectation_mc}
\end{equation}

\bibliography{kagome}

\end{document}